%% file: submission_arxiv.tex
\pdfoutput=1
\documentclass[aps,pra,10pt,twocolumn,amsmath,amssymb,notitlepage,superscriptaddress,longbibliography,nofootinbib]{revtex4-1}

\usepackage[T1]{fontenc}

\usepackage{amsmath}
\usepackage{amssymb}
\usepackage{bm}
\usepackage{graphicx}
\usepackage{xcolor}
\definecolor{darkblue}{rgb}{0,0,0.6}
\definecolor{darkred}{rgb}{0.6,0,0}
\usepackage[%
  colorlinks=true,
  urlcolor=darkblue,
  linkcolor=darkred,
  citecolor=darkblue
]{hyperref}
\usepackage{url}
\usepackage{soul}\usepackage{upgreek}

\begin{document}
\title{
Order by inertia in spinning active matter:\\ holey fluids and spin-textured crystals. 
%
%
}

\author{Camille Jorge}
\affiliation{Univ. Lyon, ENS de Lyon, Univ. Claude Bernard, CNRS, Laboratoire de Physique, F-69342, Lyon.}
\affiliation{Institute of Physics, University of Amsterdam, Science Park 904, 1098 XH, Amsterdam, The Netherlands}

\author{Denis Bartolo}
\affiliation{Univ. Lyon, ENS de Lyon, Univ. Claude Bernard, CNRS, Laboratoire de Physique, F-69342, Lyon.}
\email{denis.bartolo@ens-lyon.fr}
\begin{abstract}
Active matter sustains emergent flows at the expense of preserving structural order. The feedback between structure and viscous flows typically disrupts crystalline and liquid-crystalline organization by amplifying the very deformations they generate. Yet this destabilizing paradigm has recently been challenged by experiments showing that inertial fluid flows can stabilize few-body bound states of active spinners. 
Whether inertial active matter can sustain genuine cohesion and order at the many-body level, however, remains elusive.
Here we investigate two-dimensional assemblies of macroscopic spinners operating at high Reynolds number and uncover two phase transitions leading to the emergence of a dilute percolating fluid and a dense spin-textured crystal. 
At low density, inertial flows generate two competing interactions: anisotropic attractions and transverse Magnus forces that continuously break and reconfigure bonds. 
Together they drive a percolation transition toward a dynamically rearranging holey liquid reminiscent of the empty-liquid states observed in equilibrium patchy colloids.
At high density, the feedback between spin alignment and particle positions suppresses transverse rearrangements and yields a first-order transition toward a spin-ordered crystal. Our results demonstrate that, beyond the overdamped limit, hydrodynamic feedback can promote rather than destroy collective order, revealing a distinct regime of many-body active matter governed by inertial flows.
 \end{abstract}
\maketitle
At equilibrium, the spatial organization of matter is  immune to the details of its dynamics. 
Whether microscopic motion is inertial or viscous, thermal statistics governs structure, and phase behavior does not depend on how fast fluctuations propagate and relax. 
This protection is lost once matter is driven out of equilibrium, where structure and flow become intertwined.
In homogeneously driven soft matter, deformations generate flows that can either amplify or suppress them, leading to dynamically organized or disordered steady states in systems ranging from  sedimenting suspensions to  synthetic and living spinner crystals~\cite{crowley1976,grzybowski2000,guazzelli2011,goto2015,driscoll2017,Oppenheimer2019,Chajwa2020,massana2021,bililign2022,tan2022,shen2023,guillet2025}. 
Yet the forcing itself remains fixed: structural rearrangements and flows do not reorient gravity, the  chirality  of living cells or the magnetic field that drives the system. 

Soft active matter features a more intimate form of nonequilibrium, where actuation, flow, and structure are two-way coupled: the forces and stresses that drive the system from within are themselves shaped by the  deformations they induce.
In the viscous regime, this interplay is most often destabilizing: distortions of ordered active phases generate flows that amplify them, as strikingly illustrated by the spatiotemporal chaos of active nematics, gels and swimmer suspensions~\cite{pedley1992,saintillan2008,sanchez2012,marchetti2013,doostmohammadi2018}. 
Hydrodynamic interactions  tend to suppress long-range order in soft active matter.
Recent observations of active spinners operating at finite Reynolds number contrast with this seemingly general destabilization by revealing stable bound states sustained by inertial vortices~\cite{Chen2025,gelvan2025,shen2025}. 
Whether such inertial feedback can extend beyond isolated structures in many-body states where active matter recovers cohesion and order remains  unknown.

To determine whether inertial feedback can stabilize cohesive many-body organization, we investigate an assembly of hydrodynamically interacting spinners operating at finite Reynolds number. 
Through quantitative measurements of positional and active-spin correlations, we show that inertial forces and viscous stresses combine to produce a two-way coupling between positional and spin degrees of freedom that shapes the phase behavior. 
By systematically varying the packing fraction, we demonstrate that the system evolves through three distinct steady states separated by two phase transitions: a liquid of fractal clusters, a dynamically rearranging holey fluid, and a dense spin-textured crystal. 
Remarkably, as we show below, inertial pumping and Magnus forces generate directed interactions between spinners, providing a hydrodynamic analogue of the directional interactions that govern the phase behavior of equilibrium patchy colloids. 
At high density, however, the feedback between positional and orientational degrees of freedom collapses the holey fluid into a spinner crystal, where particle positions and spinning directions self-organize in unison. 
Beyond the specifics of our system, our results reveal how inertial hydrodynamic flows transform the destabilizing feedback typical of overdamped active matter into a mechanism of phase ordering at the many-body level.

\begin{figure*}
\includegraphics[width=\textwidth]{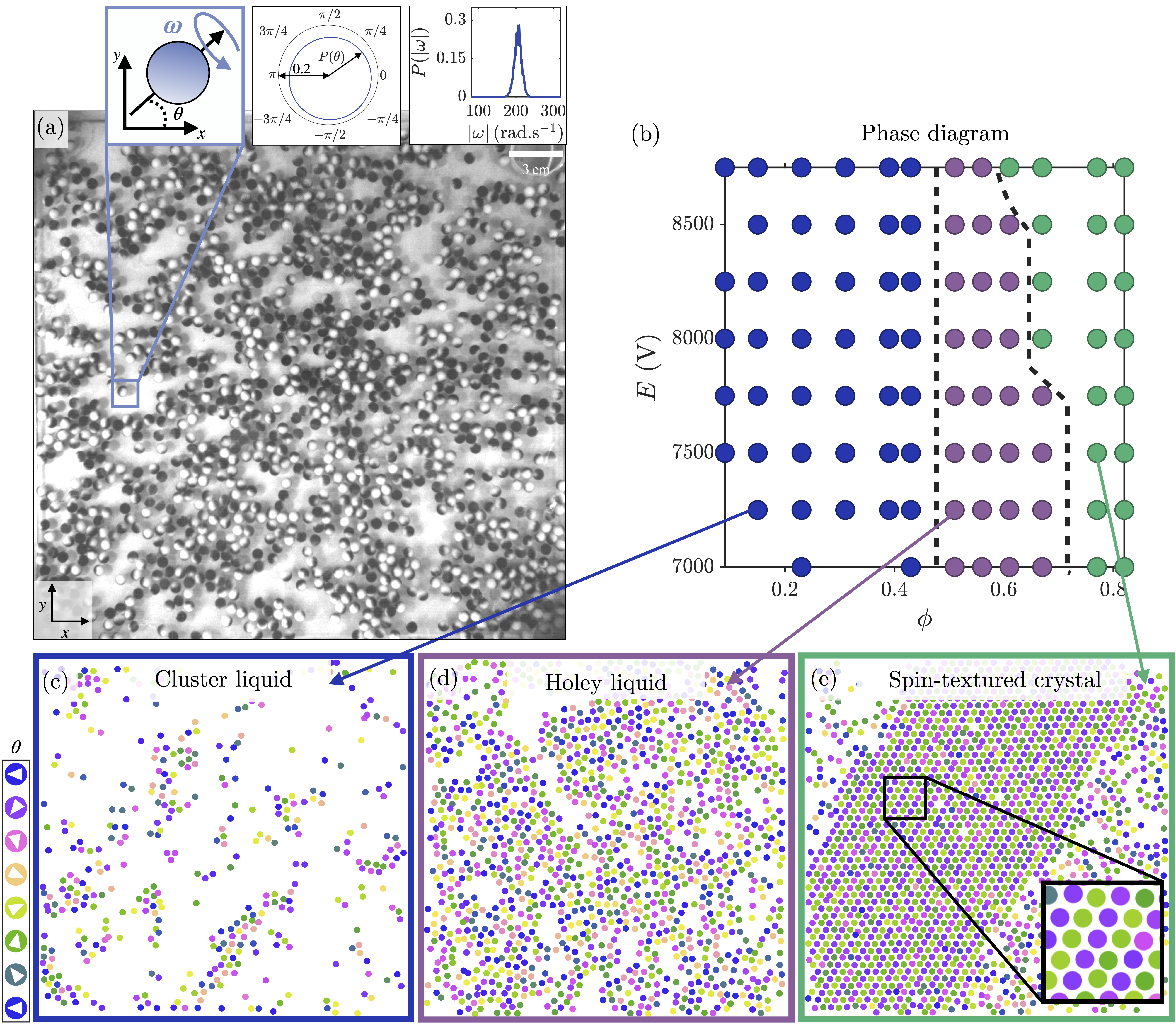}
\caption{{\bf Active spinners, cluster liquid, percolating fluid and spin-textured crystal. }
    {\bf a.} Experimental picture illustrating the spinning motion of a system of spherical plastic beads motorized using Quincke electrorotation ($\phi=0.51$). Painting the beads surfaces allows to monitor there instantaneous spinning rate $\omega$. 
    See also Supplementary Video 1.
    Left inset: scheme of a bead of spin $\boldsymbol{\omega}$ forming an angle $\theta$ with the horizontal.
    Middle inset: probability density of
    $\theta$. 
    At low packing fraction, all particles spin along random directions. ($E = 8750 V\, \rm$).
    Right inset: probability density of $ |\omega|$. 
    The rotation speed is constant ($E = 8750 \, \rm V$ and $\phi = 0.39$).
    {\bf b.} Phase diagram constructed from 97 different experiments. The three phases are mostly determined by the spinners packing fraction. The phase boundaries hardly depend on the magnitude of the voltage $\mathbf E$.
    {\bf c.}, {\bf d.} and {\bf e.}  Snapshots of three instantaneous spinner configurations recorded at three different packing fractions. 
    The particles are colored according to their instantaneous orientation.
    See also Supplementary Videos 2, 3 and 4.
    {\bf c.} $\phi=0.15$:
    The spinners form an isotropic liquid of clusters. 
    {\bf d.} 
    $\phi=0.51$:
    The spinners form an isotropic percolating  fluid.
    Unlike  in colloidal gel their structure is highly dynamical ~\cite{royall2021,sciortino2011}.
    {\bf e.} $\phi=0.67$:
    The active particles phase separate into a spin-textured crystals where spins form ferromagnetic lines along a principal axis of the crystal and alternate orientation from one line to the next. 
    The crystal coexists with a dilute cluster liquid.
    }
    \label{Fig1}
\end{figure*}
\section*{\bf Active-spinner experiments.} We detail the experimental setup showed in Figure~\ref{Fig1}a in SI. 
In brief, we confine plastic beads of radius $a = 3\,\mathrm{mm}$ immersed in a hexadecane solution between two transparent electrodes separated by $8.5\,\mathrm{mm}$ and apply a DC electric field along $\hat{\mathbf{z}}$ to activate Quincke electro-rotation~\cite{quincke1896ueber} (Figure~\ref{Fig1}a). 
As in colloidal-roller experiments~\cite{Bricard2013}, once the voltage exceeds the Quincke threshold $E_Q = 6000\,\mathrm{V}$ the beads roll along the electrodes (see SI). 
However, increasing the voltage further ($E \geq 7000\,\mathrm{V}$) lifts the particles off the surface~\cite{pradillo2019}, suppresses their translational motion, and produces a hovering monolayer of active spinners (Figure~\ref{Fig1}, Supplementary Videos 1-4 and SI).
In the following, we  focus on this spinning regime and track the active particle positions $\mathbf{r}_i(t)$, and rotation vectors $\bm{\omega}_i(t)$, which we refer to as the particle spin (Figure~\ref{Fig1}a). 
Unlike in most active spinner experiments and simulations,  the spins are not imposed, but instead evolve dynamically through particle interactions~\cite{Bricard2013,Braun2026}. 
Moreover, the $\bm \omega_i$ are not Ising-like variables restricted to a fixed $z$ axis. 
In our 90 experiments, illustrated in Figure~\ref{Fig1}, each particle can rotate about an arbitrary axis in the $xy$-plane at an approximately constant rate.
  Their magnitude $\omega_i$ is narrowly distributed  (Fig.~\ref{Fig1}a) and varies  between $24\,\rm Hz$ and $34\,\rm Hz$ as we increases $E$. 
  As a result, the induced flows are associated to Reynolds number $Re = a^2 \omega / \nu$ ranging from 
    $300$ to $400$, 
  revealing that the spinners operate in an inertia-dominated regime ($\nu=4.5\,\rm cSt$ is the kinematic viscosity of the fluid). 
\\

\subsection*{\bf From cluster liquids to spin-textured crystals.} Figure~\ref{Fig1}d and Supplementary video 2 shows that dilute ensembles of interacting spinners form an isotropic liquid phase. 
Even at low packing fraction ($\phi < \phi_p =0.43$), small clusters continuously assemble and disassemble, indicating effective attractive interactions that cannot classically originate from motility~\cite{buttinoni2013,cates2015,ginot2018}, as the spinners exhibit nearly no self-propulsion. 
At intermediate packing fractions ($\phi_p < \phi < \phi_x = 0.67 \pm 0.7$), the fractal clusters undergoes a sharp geometrical transition and self-organizes into a holey fluid in which most particles belong to a single percolating cluster (Figure~\ref{Fig1}d).
Unlike colloidal gels~\cite{zaccarelli2007,royall2021}, this holey fluid is not dynamically arrested but continuously rearranges its geometry, as seen in Supplementary Video 3 and quantified in SI. 
At $\phi = \phi_x$, the translational and spin-rotational symmetries of the holey fluid are both spontaneously broken. 
The holey dynamical structure collapses into a dense hexagonal crystal that coexists with a dilute cluster liquid (Figure~\ref{Fig1}e and Supplementary Video 4). 
Simultaneously, the spins organize into system-spanning ferromagnetic lines whose orientation alternates from one line to the next. 
We now demonstrate that these three nonequilibrium steady states define a  phase diagram where two phase transitions are chiefly controlled by the spinner packing fraction $\phi$ (Fig.~\ref{Fig1}e).

\begin{figure*}
\includegraphics[width=\textwidth]{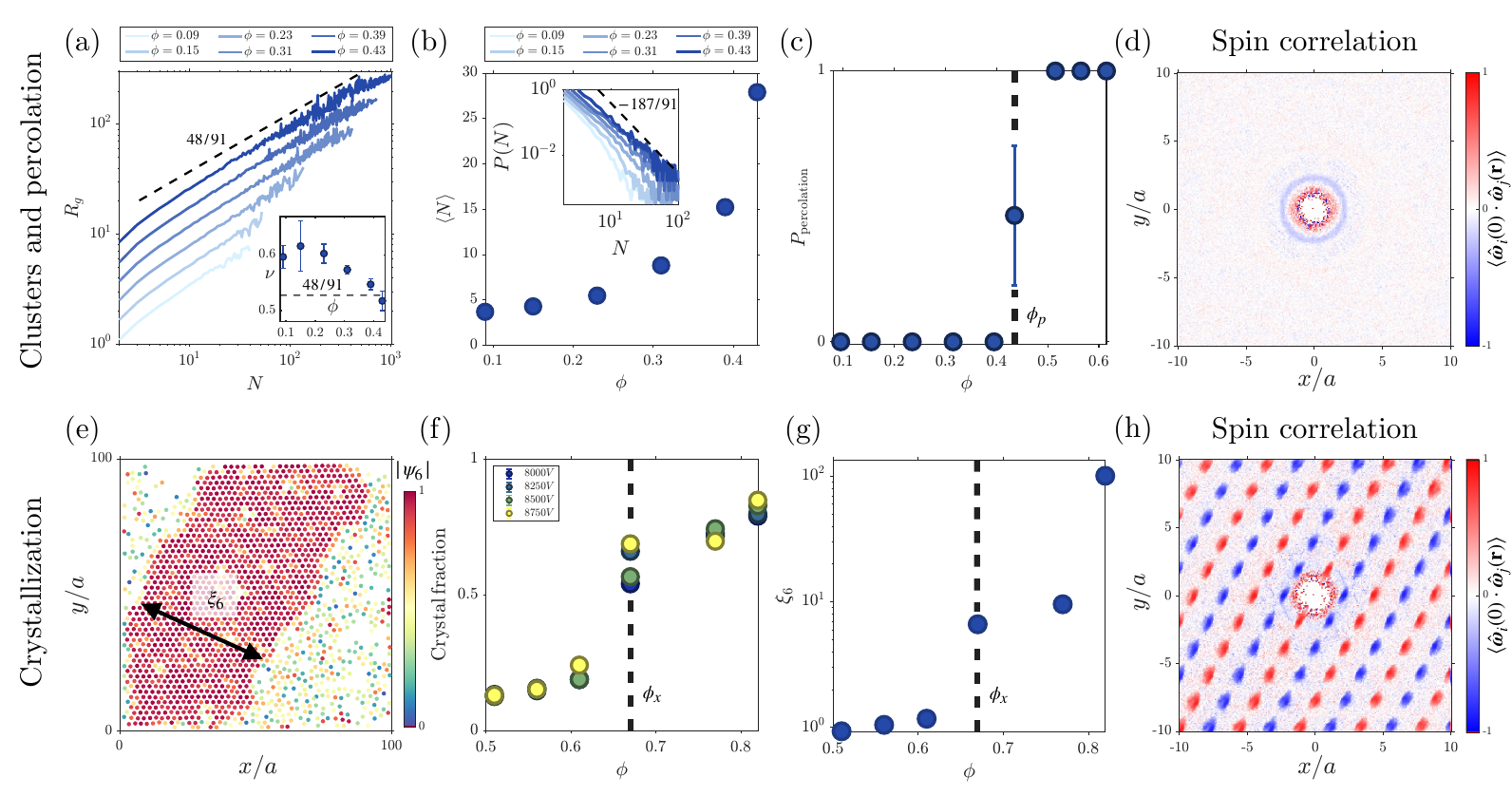}
\caption{{\bf Percolation transition, phase separation and spin ordering. }
    {\bf a.} The gyration radius of the cluster grow as a power law $R_{\rm g}\sim N^\nu$ at all packing fractions (all $E$ averaged).
    Inset: Exponent $\nu$ plotted versus $\phi$. The dotted line corresponds to the  exponent $\nu=48/91$ computed at the classical percolation transition, see e.g.~\cite{duplantier1987exact,nienhuis1984}. 
    {\bf b.} The average number of spinners in a cluster $\langle N \rangle$ grows with $\phi$ (average over all $E$ values). The variations blow up as $\phi$ approaches $\phi_p=0.43$. 
    Inset: the sizes  of the clusters are power-law distributed. The dashed line corresponds to the power law expected at the percolation transition $\tau=187/91$~\cite{duplantier1987exact,nienhuis1984} (average over all $E$ values). 
    {\bf c.} The probability to observe a percolating cluster in an experiment jumps from 0 to 1 at $\phi=\phi_p=0.43$ ($E=8250 \, \rm V$).
    {\bf d.}  Typical spin-spin correlations in the holey liquid phase ($E=7750 \, \rm V$, $\phi=$). 
    The spinners only feature very short range isotropic correlations. No long-range spin order builds up.
    {\bf e.} Snapshot of an experiments showing the coexistence between a crystal and a liquid phase.
    The particles are colored according to the value of their bond orientational order (see SI).
    {\bf f.} The area fraction spanned by the crystal phase increases with $\phi$ and discontinuously jumps at $\phi=\phi_x=0.67$.
    {\bf g.}  The correlation length $\xi_6$ of the bond-order parameter $\phi_6$ measure the extent of the crystals.
    $\xi_6$ increases sharply at $\phi_x$.
    {\bf h.} Spin-spin correlations in the crystal phase. 
    Spins are aligned in the axial direction and anti-aligned from one line to the next. 
    This alternate order persists over system spanning scales ($E=8750 \, \rm V$, $\phi=0.67$).
}
    \label{Fig2}
\end{figure*} 

\section*{Percolation, phase separation and spin ordering}
We begin by showing that the cluster liquid undergoes a percolation transition, leading to a dynamically rearranging state reminiscent of the empty liquids observed in equilibrium patchy colloids~\cite{Bianchi2006}.
Figures~\ref{Fig2}a,b show that the mean cluster size increases with the packing fraction $\phi$. 
However, these clusters are not compact droplets of well-defined size. 
Instead, they form fractal objects characterized by a power-law size distribution (Figures~\ref{Fig2}a,b). 
We find that the probability $P$ of observing a system-spanning cluster jumps from zero to one at $\phi_p=0.43$ in Figure~\ref{Fig2}d. 
Moreover, at $\phi_p$ the gyration-radius and cluster-size exponents match those expected at the percolation threshold ($\nu = 91/48$ and $\tau = 187/91$)~\cite{stauffer1979,nienhuis1984,saleur1987}. 
Together, these results establish that the cluster liquid and the holey fluid are two distinct nonequilibrium phases.
Despite this geometrical transition, both phases remain isotropic and lack any form of spin order. 
In each case, the rotation vectors $\omega_i$ display neither macroscopic alignments nor long-range correlations (Fig.~\ref{Fig2}d).

The collapse of the percolating network into a spin-textured crystal belongs to a fundamentally different class of transition. 
Figure~\ref{Fig2}e shows that a crystalline phase coexists with a dilute liquid in which the bond-orientational order parameter $\psi_6$ remains vanishingly small. 
Furthermore, Figures.~\ref{Fig2}f and~\ref{Fig2}g show that both the crystalline area fraction and, equivalently, the correlation length $\xi_6$ of the $\psi_6$ field undergo discontinuous variations at $\phi_x=0.67$. 
Together, these observations indicate that crystallization results from a phase-separation process, consistent with a first-order transition.
Crucially, crystallization coincides with the emergence of long-range spin order throughout the crystal (Fig.~\ref{Fig2}h). 
This simultaneous breaking of translational and spin-rotational symmetries suggests a mutual dependence between positional interactions and spin torques, which we now quantify experimentally.\\

\begin{figure*}
\includegraphics[width=\textwidth]{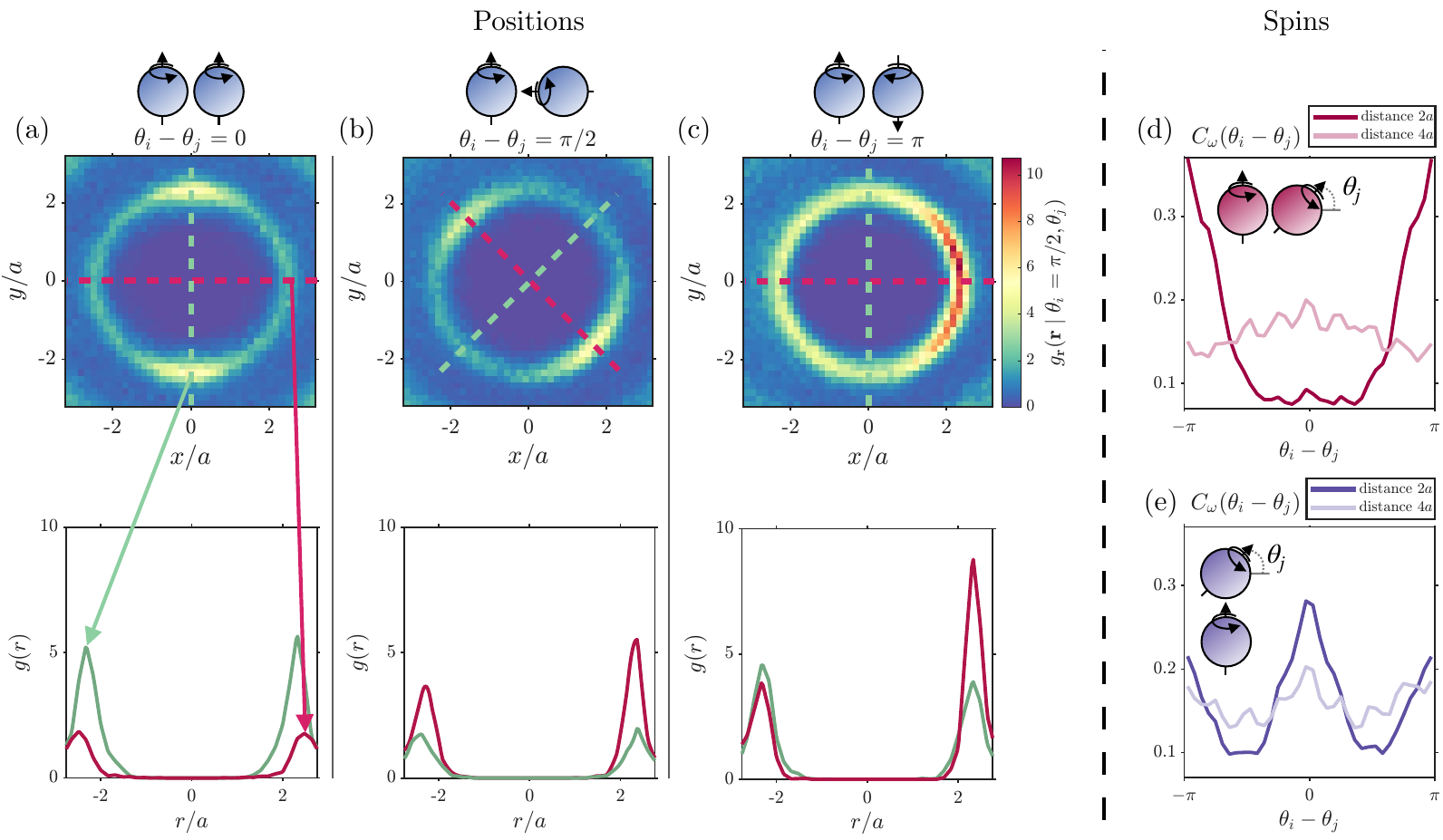}
\caption{{\bf From correlations to interactions. }
    Spin-resolved two-point distributions
    measured for different spin configurations ($g_{\bf r}(\mathbf r, \theta_1,\theta_2)$).
    {\bf a.} When the two  spins  $\omega_1$ and $\omega_2$ are aligned,t he probability to find a second spinner is maximal in the axial direction. 
    {\bf b.} When the two  spins are orthogonal, the probability is maximal in the lower right and upper left quadrants.
    {\bf c.} When, the two spins are anti-parallel, $g_{\bf r}$ features as strong peak at contact in all directions, but with pronounced the left-right asymmetry.
    {\bf d., e. and f.} Directional cross-sections of maps {\bf a.}, {\bf b.} and {\bf c.}) along the red and green dashed lines.
    {\bf g.} 
    Position-resolved angular pair-correlation function for two spins sitting on  the $x-$axis ($C_{\bm \omega}$). The first spin $\bm \omega_1$ points along the $y-$axis.
    The most probable configuration corresponds to anti-aligned spins at short distances.
    At distances larger than $5a$, we measure  no spin correlations.
    {\bf h.} Same quantities measured when the spinners sit on the $y$-axis. 
    At short distance, the most probable configuration corresponds to spins pointing in the same direction. 
    At long distance, the spin orientations are uncorrelated.
}
    \label{Fig3}
\end{figure*}
\section*{\bf Symmetries of the two-spinner interactions.}
\subsection*{Pair correltaions}
To characterize the symmetries of the  pair interactions governing the spinner dynamics, we analyze our  experiments  in the dilute regime ($\phi<0.43$), and focus on two central observables. 
First, we generalize the standard pair-correlation function by conditioning it on the  spin orientations of the interacting particles. 
Specifically, we introduce the spin-resolved two-point distribution $
g_{\mathbf r}(\mathbf r \mid \theta_1,\theta_2)
= P(\mathbf r_2=\mathbf r \mid \mathbf r_1=0, \theta_1, \theta_2)$,
where $\theta_i$ denotes the in-plane orientation of the spin vector $\bm\omega_i$ (see Fig.~\ref{Fig1}a). 
Measurements are performed in the reference frame where $\mathbf r_1=0$ and  $\bm\omega_1\propto\bm{\hat{y}}$.
We first consider parallel spin configurations. 
Figure~\ref{Fig3}a shows that the pair correlation is enhanced along the spin axis, indicating a pronounced axial attraction. 
When the spins are orthogonal (Figure~\ref{Fig3}b), the correlations are again strongly anisotropic, but their maxima are localized in the upper-right and lower-left quadrants  consistent with a spin-dependent transverse  attraction. 
Lastly, for antiparallel spins ($\theta_1=\theta_2+\pi$), the particles are likely to be in contact ($r=2a$) in all directions  yet the correlations exhibit a stronger  left–right asymmetry (Figure~\ref{Fig3}c). 
The asymmetries in the correlations plotted in Figures.~\ref{Fig3}b,c directly reveals a non-reciprocal component of the effective interactions.

We now infer the symmetry of the effective spin interactions, through the angular pair-correlation function. 
Working in the reference frame where particle 1 is at the origin ($\mathbf r_1=0$), and $\boldsymbol{\omega}_1 \parallel \hat{\mathbf y}$, we  measure the probability that the two particles differ in spin orientation by an angle $\Delta\theta$ when they are separated by a vector $\mathbf r$:
$
C_{\bm \omega}(\Delta\theta \mid \mathbf r)
= P(\theta_i-\theta_j=\Delta\theta \mid \mathbf r_2-\mathbf r_1=\mathbf r)
$.  
When the second particle lies in the direction transverse to $\bm{\omega}_1$, $C_{\bm \omega}$ peaks at $\Delta\theta=\pi$ (Figure~\ref{Fig3}d), which indicates a preference for antiparallel alignment, and thus effective antiferromagnetic interactions in the transverse direction. 
Conversely, when the separation is axial, (Figure~\ref{Fig3}e), $g_{\bm \omega}$ is maximal at $\Delta\theta=0$, which echoes ferromagnetic alignment along the spin axis. 
In both cases, the correlations decay rapidly with increasing $r$, signaling that the effective spin interactions are short-ranged.

The pair correlations $g_{\mathbf r}$ and $C_{\boldsymbol{\omega}}$ constrain the symmetry of the effective interactions, and reveal three essential features: spin-dependent directional attraction, position-dependent spin alignment, and non-reciprocity. 
Such properties cannot result from the reciprocal repulsive forces between Quincke electrostatic dipoles ~\cite{Bricard2013,Taylor1969}. 
They instead point toward  dominant hydrodynamic couplings.

\begin{figure*}
\includegraphics[width=\textwidth]{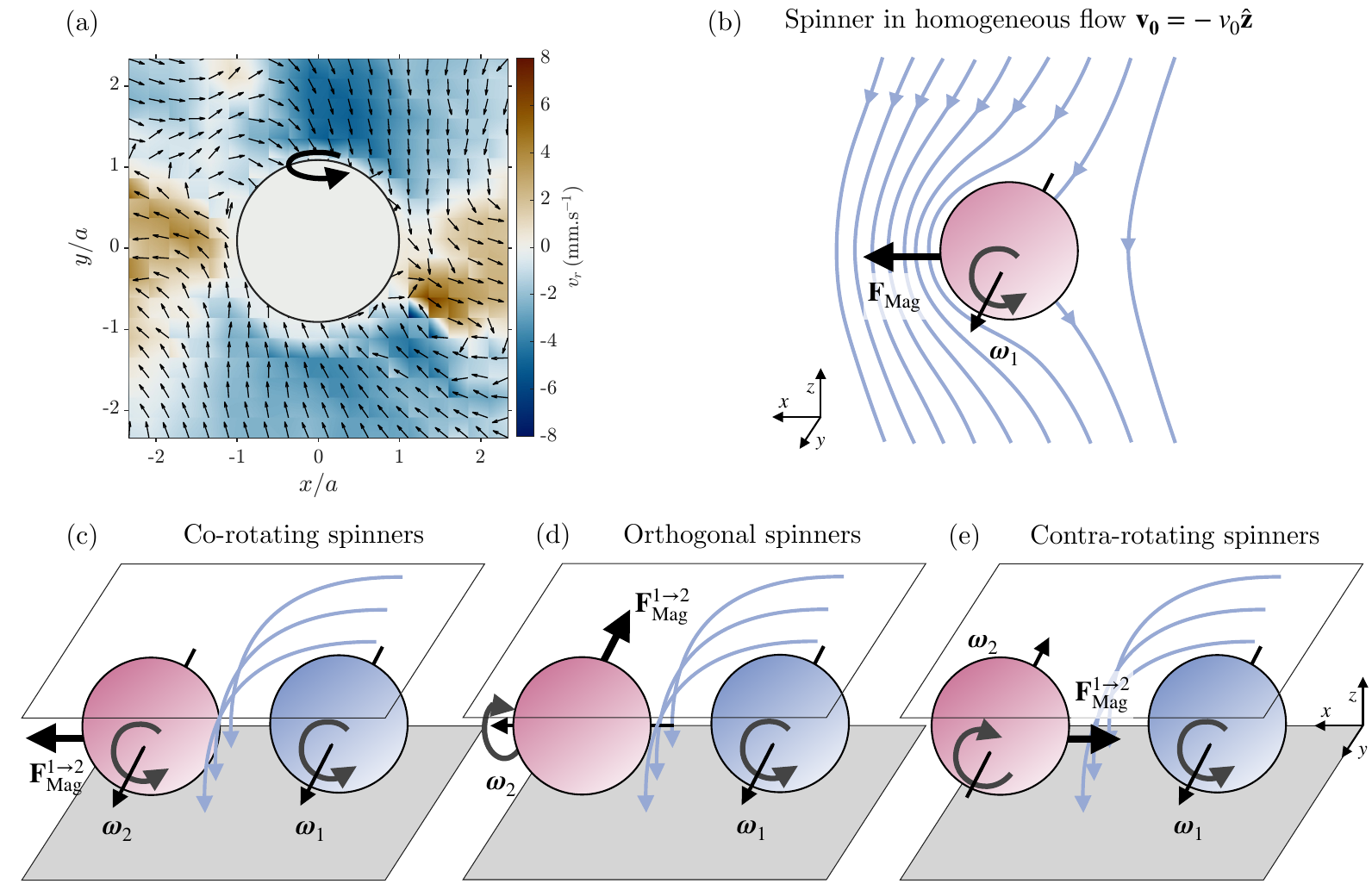}
\caption{{\bf Inertial pumping, Magnus forces and (anti)aligning torques. }
    {\bf a.} Velocity field measured by Particle Imaging Velocimetry of the fluid flow averaged over the $z$-direction around a spinner ($E = 8500 V\, \rm$). 
    Integrating over $z$, the main vortex contributions vanish revealing the secondary flow field. 
    Centrifugal force expels the fluid in the transverse direction and mass conservation implies axial pumping. 
    The left-right asymmetry is a consequence of the residual spinning motion. See SI for experimental details and for a detailed characterization. 
    {\bf b.} Illustration of the Magnus effet acting on a spinner in a homogeneous flow. Due to the spinner's rotation, the flow is accelerated at the left the spinner and slowed down at the right, creating a net force to the left.
    {\bf c.} Resulting interaction between two co-rotating spinners. The resulting force of $2$ on $1$ is repulsive.
    {\bf d.} Resulting interaction between two orthogonal spinners. The resulting force of $2$ on $1$ is transverse.
    {\bf e.} Resulting interaction between two contra-rotating spinners. The resulting force of $2$ on $1$ is attractive.
}
    \label{Fig4}
\end{figure*}
\subsection*{\bf Inertial hydrodynamic interactions.}
To elucidate how hydrodynamics couples the internal and positional degrees of freedom, we recall that the Reynolds number associated with the vortex generated by a spinner exceeds 300, clearly placing the system in an inertial regime. 
Each spinner behaves as a localized vortical source, or ``vortlet''~\cite{Chen2025}, dressed with a  secondary flow with a strong quadrupolar structure revealed by Figure~\ref{Fig4}a and SI, see also~\cite{bickley1938lxv,goto2015,Chen2025,shen2025}. 
Centrifugal inertia and mass conservation generate transverse fluid expulsion and axial pumping around each spinner.
This quadrupolar flow results in a strong axial advection towards the spinner, consistent with the pair correlations reported in Figure~\ref{Fig3}b. 
However, this inertial pumping is independent of the spin $\boldsymbol{\omega}_2$ of the second particle, and therefore cannot account for the full spin-dependent structure of the correlations $g_{\mathbf r}$ (Figures.~\ref{Fig3}a,c).
A crucial observation is that the  particles are not  merely advected by the flows. 
They are not a passive tracer, but actively spin while being advected by the flow generated by their neighbors.
Consequently, and as illustrated in  Figure~\ref{Fig4}b, a particle  with a spin $\bm \omega$ experiences a transverse lift, a Magnus force, $\mathbf F_{\rm Mag}=C\rho\mathbf v(\mathbf r) \times \bm\omega $ when the fluid of density $\rho$ is locally moving  at a velocity  $\mathbf v(\mathbf r)$ ($C$ a geometrical factor of order $a^3$)~\cite{Rubinow1961,guyon2015}. 
This phenomenon leads to counterintuitive hydrodynamic interactions.
For separations smaller than the electrode spacing, and ignoring  secondary-flow corrections, we can approximate the vortex generated by a spinner rotating with $\boldsymbol{\omega}_1$ as $
\mathbf v_1(\mathbf r) =(a/r)^3\boldsymbol{\omega}_1 \times \hat{\mathbf r}$ 
(at larger distances the vortex is exponentially attenuated by confinement)~\cite{gelvan2025}. 
Substituting this expression into the Magnus force  yields the interaction force
\begin{equation}
\mathbf F_{\rm Mag}^{1\to2}
= C\left(\frac {a} {r_{12}}\right)^3 \rho
(\boldsymbol{\omega}_1 \times \hat{\mathbf r}_{12}) \times \boldsymbol{\omega}_2 .
\label{Eq:Magnus}
\end{equation}
This formula immediately reveals three key properties. 
First, Magnus interactions  depend on both particle spins, demonstrating explicitly that the positional dynamics is directly coupled to the internal spin degrees of freedom. 
Second, these interactions feature an action–reaction symmetry only when the two spins are collinear as
$\mathbf F_{\rm Mag}^{1\to2}+\mathbf F_{\rm Mag}^{2\to1}
\propto
(\boldsymbol{\omega}_1 \times \boldsymbol{\omega}_2)\times \mathbf r_{12}$. 
Finally, when the spins are orthogonal, the Magnus forces either vanish or become purely transverse to the separation vector $\mathbf r_{12}$. 

These remarkable properties provide a natural interpretation of the measured pair correlations from the joint action of inertial pumping and Magnus force (Figure~\ref{Fig3}). 
When the spins are parallel, Eq.~(\ref{Eq:Magnus}) predicts that the Magnus force is purely  repulsive along the direction normal to the spins (Figure~\ref{Fig4}c). 
Combined with the axial pumping induced by the inertial vortical flow, this mechanism  accounts for the strong anisotropic structure observed in Figure~\ref{Fig3}a.
When the spins are orthogonal, Eq.~(\ref{Eq:Magnus}) predicts forces that are either vanishing or purely transverse (Figure~\ref{Fig4}d). 
A second particle can approach a spinner only when it lies in its upper-right or lower-left quadrants. Combined with contact repulsion this interaction naturally explains the tilted correlations reported in Figure~\ref{Fig3}b.
When the spins are antiparallel, the Magnus forces act with opposite signs on opposite sides of the reference particle (Figure~\ref{Fig4}e). 
Combined with inertial pumping, they produce a net attraction in all directions, consistent with the correlations observed in Figure~\ref{Fig3}c, and for the spinner pairing reported in~\cite{Chen2025,gelvan2025}. 
The left–right asymmetry of the pair correlations then arises from the residual rolling motion of the hovering spinners.

Having established how particle spins shape the effective interparticle forces, we now turn to the reciprocal problem: how the torques acting on the spins depend on particle positions. 
Elucidating this mechanism closes the feedback loop that couples the rotational and positional degrees of freedom of the spinners.
To understand how spins interact, it is useful to recall that Quincke rotation originates from an instability that spontaneously breaks the rotational symmetry of the electro-hydrodynamic problem. 
As a result, an isolated Quincke particle can rotate around any axis lying in the plane perpendicular to the applied electric field~\cite{Taylor1969}. 
An external torque explicitly breaks this symmetry and selects the stationary rotation axis~\cite{Braun2026}, in direct analogy with the alignment of an $XY$ spin in a magnetic field.
Consider a spinner located at the origin and rotating with angular velocity $\boldsymbol{\omega}_1=\omega\,\hat{\mathbf y}$. 
At short distance, the lubrication flow it generates induces a hydrodynamic torque on a second spinner located at $\mathbf r$:
$\mathbf T_2=\left(X^C \mathbf P+Y^C \mathbf P^\perp\right)\cdot \bm \omega_1$, where $X^C>0$ and $Y^C<0$ are two constants and $\mathbf P$ and $\mathbf P^\perp$ are the projection operators on the $\mathbf r$ and $\mathbf r^\perp$ directions, see SI for details and~\cite{jeffrey1984}. 
This torque sets the direction of $\boldsymbol{\omega}_2$, and its direction depends on the relative position of the two particles. When the spinners are aligned along the rotation axis, the shear flow between them favors co-rotation, resulting in a ferromagnetic-like coupling.
By contrast, when the particles are separated along the $x$-axis, rolling interactions orient $\boldsymbol{\omega}_2$ opposite to $\boldsymbol{\omega}_1$ (see SI).
These hydrodynamic interactions therefore produce position-dependent spin couplings that directly account for the symmetry of the spin–spin correlation functions $C_{\bm \omega}$ reported in Figure~~\ref{Fig3}d. 
The hydrodynamic feedback loop is closed.

\section*{\bf From pair interactions to phase behavior}
Having established the structure and symmetries of pair interactions, we can now rationalize the phase behavior reported in Figures~\ref{Fig1} and~\ref{Fig2}. 
At low packing fractions the spatial organization of the suspension results from the competition between two nonequilibrium mechanisms. 
Inertial pumping produces a strongly anisotropic attraction along the spin axis. 
They effectively endow the particles with patch-like interactions that promote the formation of directed bonds (Figure \ref{Fig4}a), see also~\cite{gondret1996}. 
In contrast, the non-reciprocal transverse Magnus forces, together with residual rolling motion, tend to disrupt these axial bonds, or to reorient them when the spin directions differ (Figures~\ref{Fig3}b,c and~\ref{Fig4}b).
This competition is reminiscent of the physics of equilibrium patchy colloids, where directional attractions compete with thermal fluctuations~\cite{zaccarelli2007,sciortino2011,russo2022}. 
In those systems the finite valence of the interactions suppresses bulk phase separation and instead stabilizes percolating fluid or empty liquid phases. 
Although our spinners are neither Brownian nor in equilibrium, transverse motions induced by Magnus forces play a role analogous to thermal noise by continuously breaking and rearranging directed bonds.
As a result, inertial attraction promotes the formation of clusters. At the same time, the strong anisotropy of the interactions limits the effective valence of the spinner attraction, and prevents bulk phase separation and condensation into compact drops before connectivity percolates at system-spanning scales.

Unlike at low packing fraction, the collapse of the active percolating fluid into a spin-textured crystal above $\phi_x$ relies both on spin-dependent forces and on position-dependent spin interactions. 
At high density the feedback between particle positions and spin orientations becomes strong enough to arrest the spin dynamics. Instead of continuously reorienting, the spins freeze into ferro–antiferro configurations along the axial and transverse directions.
In these configurations the  Magnus forces that previously disrupted directed bonds vanish by symmetry. 
As a result, they no longer generate transverse rearrangements and instead contribute purely axial and lateral attractions between neighboring spinners (Figure~\ref{Fig3}c). 
The effective valence of the interactions therefore increases dramatically, and no dynamical mechanism remains capable of breaking the bonds between particles.
Under these conditions phase separation becomes unavoidable: the percolating active fluid collapses into dense phase of monodisperse spinners whose cohesion is reinforced by the emergent spin order, which simultaneously freezes the positional and orientational degrees of freedom. 
The resulting crystal can coexist with a dilute and isotropic liquid phase composed of finite clusters.\\

\section*{\bf Conclusion and outlook.}
Our experiments reveal a regime of active matter in which  inertial hydrodynamics, rather than dissipation alone governs collective organization. 
They highlight how feedback between flows and active degrees of freedom can control phase transitions and structural order, here through an effective patchiness leading to percolating fluids and spin-textured crystals.
Looking ahead, exploring stronger driving conditions may uncover even richer dynamics, where active particles and inertial turbulence  interact and organize one another.

\section*{Acknowledgments} We thank W. Irvine, J. Lintuvuori, A. Snezhko and N. Oppenheimer for insightful feedbacks and discussions.  
 This project  received funding from the European Research Council (ERC) under the European Union’s Horizon 2020 research and innovation program (grant agreement No. [101019141]) (DB).

%

\clearpage
\appendix

\setcounter{equation}{0}
\setcounter{figure}{0}
\setcounter{table}{0}
\setcounter{page}{1}
\setcounter{section}{0}

\renewcommand{\theequation}{S\arabic{equation}}
\renewcommand{\thefigure}{S\arabic{figure}}
\renewcommand{\thetable}{S\arabic{table}}
\renewcommand{\thesection}{S\Roman{section}}
\renewcommand{\thepage}{S\arabic{page}}


\clearpage
\onecolumngrid 

\begin{center}
    \vspace*{1cm}
    {\large\bf Supplemental Material for:} \\   
    \vspace*{0.2cm}
    {\Large\bf Order by inertia in spinning active matter:\\ holey fluids and spin-textured crystals. } \\ 
    \vspace*{0.6cm}
    {\large Camille Jorge$^{1,2}$ and Denis Bartolo$^{1}$} \\ 
    \vspace*{0.3cm}
    {\small\it $^1$Univ. Lyon, ENS de Lyon, Univ. Claude Bernard, CNRS, Laboratoire de Physique, F-69342, Lyon.}\\ 
    {\small\it $^2$Institute of Physics, University of Amsterdam, Science Park 904, 1098 XH, Amsterdam, The Netherlands} \\ 
    \vspace*{1cm}
\end{center}

\twocolumngrid

\input{SI_content.tex} 

\end{document}

%% file: SI_content.tex
\title{
 Supplementary Information
}

\author{Camille Jorge}
\affiliation{Univ. Lyon, ENS de Lyon, Univ. Claude Bernard, CNRS, Laboratoire de Physique, F-69342, Lyon.}
\affiliation{Institute of Physics, University of Amsterdam, Science Park 904, 1098 XH, Amsterdam, The Netherlands}
\author{Denis Bartolo}
\affiliation{Univ. Lyon, ENS de Lyon, Univ. Claude Bernard, CNRS, Laboratoire de Physique, F-69342, Lyon.}

\maketitle

\renewcommand{\theequation}{S\arabic{equation}}
\renewcommand{\thefigure}{S\arabic{figure}}

\section{Methods}

\subsection{Experimental setups}
\subsubsection{Interacting spinners}

\begin{figure*}[ht]
 \includegraphics[width=\linewidth]{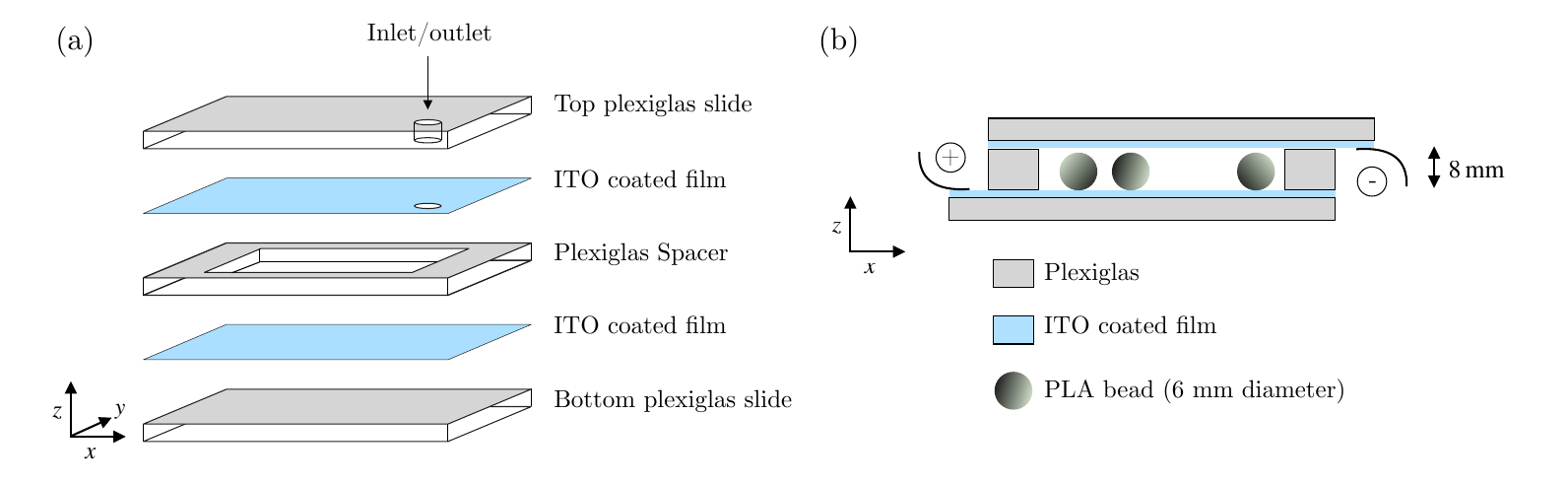}
     \caption{
     (a) Exploded view  of the experimental setup. The size of the top and bottom plexiglas slides is $30 \times 30 \rm \, cm^2$. The inner side of the spacer is $28 \times 29 \rm \, cm^2$.
     (b) Side view. Side view of the experimental setup. The electrodes protrude slightly beyond the edges so that electrical wires can be attached to them.
     }
    \label{device}
\end{figure*}

The experimental setup is very similar to the standard Quincke-roller experiments described in detail, for example in \cite{chardac2021topology}. The characteristic length scales, however, differ by three orders of magnitude. 
Here, the spinners are $6\,\mathrm{mm}$ PLA beads (polylactic acid) immersed in a solution of hexadecane and AOT salt (dioctyl sulfosuccinate sodium salt) at a concentration of $5.5\times10^{-3}\,\mathrm{wt\%}$. 
A schematic of the device is shown in Fig.~\ref{device}. 
It consists of two transparent PET films coated with indium tin oxide (ITO, Sigma) and taped onto thick PMMA plates. 
The two plates are bonded to an $8\,\mathrm{mm}$-thick PMMA frame. 
The frame defines the accessible area for the particles, with dimensions $28 \times 29\,\mathrm{cm^2}$. A $3\,\mathrm{cm}$ hole is drilled in the upper PMMA plate to load the particles and the hexadecane solution.

We drive the rotation of the beads by taking advantage of Quincke electrorotation~\cite{quincke1896ueberSI,Taylor1969SI}. 
A DC electric field is applied across the two ITO electrodes using a low-frequency generator coupled to a voltage amplifier. The applied voltage ranges from $6000$ to $9000\,\mathrm{V}$.
In order to track both the position and the rotation of the beads, we coat approximately half of the surface of each bead with black spray (see Fig.~\ref{detection}a and Supplementary Video 1).  

\subsubsection{Single spinner experiments}
Measurements of the flows generated by the spinners or of their elevation as a function of the electric field require isolating a single spinner and recording it in close-up. We detail our experimental methods in this section.

The experimental setup is a scaled-down version of that described above (dimensions $5 \times 7.5 , \rm cm^2$).
To vizualize the dynamics in the $xz-$plane, we replace one of the PMMA plates forming the frame with a glass plate.
We trap a single bead  before it is hermetically sealed. 

We use paramagnetic colloids of $20 \, \rm \mu m$ diameter  dispersed in the solvent to visualize the flow generated by the bead motions. 
We do not exploit the paramagnetic properties of the colloids; we use them because they are not subject to the Quincke effect in hexadecane and therefore do not roll on the electrodes.


\subsection{Image Acquisition and data analysis}

\subsubsection{Image acquisition}
 We illuminate the system using standard desk lamps and adjust the lighting angle to minimize specular reflections as much as possible. 
The experiments are recorded with a LUX160 (Ximea) camera equipped with a zoom lens.  
To ensure sufficient temporal resolution when measuring the rotation of the beads, we record images at a frame rate of $300\,\mathrm{Hz}$ for $10\,\mathrm{s}$. 
This frame rate is approximately $25$–$30$ times larger than the bead rotation frequency. 
Once the electric field is switched on, we wait $5\,\mathrm{min}$ for the system to reach a steady state before starting the acquisition.

\subsubsection{Particle detection and tracking}

\begin{figure*}[ht]
 \includegraphics[width=\linewidth]{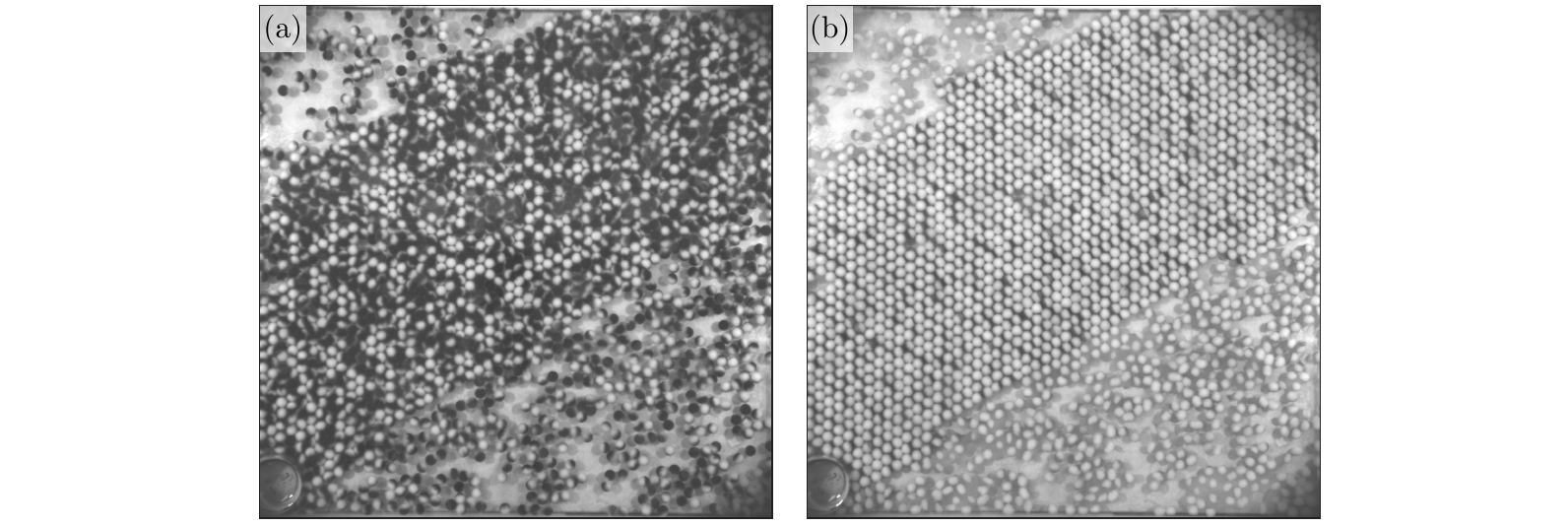}
     \caption{\textbf{Image processing prior to detection.}
     (a) Raw experimental image.
     (b) Post processed image. The maximum-intensity projection transforms the rotating black and white beads into nearly circular white spots than can be detected using standard methods.}
    \label{detection}
\end{figure*}

Since the particles are bicolored, we need to adapt the standard methods used to detect the positions of homogeneous spheres (Fig.~\ref{detection}a). 
First, we compute the maximum-intensity projection over a number of frames corresponding approximately to the rotation period of a spinner. 
Because the particles translate very little during this time interval, the resulting image consists of a collection of nearly circular homogeneous spots (Fig.~\ref{detection}b). 
We then detect their centers using a standard Gaussian fitting algorithm \cite{blair2008matlab}. 
Finally, we track the particle trajectories using the Crocker and Grier algorithm~\cite{crocker1996methods}.

\subsubsection{Rotation axis and rotation speed}

\noindent {\bf Spinning axis.}
To measure the orientation of the spinning axis of the beads, we first perform a standard particle image velocimetry (PIV) analysis using PIVlab on successive images~\cite{thielicke2021particle}. 
We then apply a circular mask centered on each particle to measure the velocity field only at the particle surface. 
Next, we compute the spatial average of the surface velocity for each particle, average it over one rotation period, and normalize it. 
Finally, applying a $\pi/2$ rotation to this vector defines the orientation of the spin $\bm{\omega}_i$.\\

\noindent{\bf Rotation speed.}
Instead of relying on PIV measurements, we determine the rotation speed of each spinner using a spectral method.
We compute the temporal Fourier transform of the image intensity and spatially average it over the surface of each particle. 
The resulting spectrum exhibits a peak at the spinning frequency due to the strong contrast between the black and white sides of the bicolored beads.\\

%
\subsubsection{Height and flows of an isolated spinner }
\noindent\textbf{Height measurement.}
We image a single bead from the side using a LUX160 (Ximea) camera mounted on a Nikon AZ100 microscope. 
The image size is $1384 \times 180\,\mathrm{px^2}$ and the resolution is $24 \,\rm px /mm$. 
We detect the particle center using the circle detection algorithm \texttt{viscircles} implemented in MATLAB and measure its distance from the bottom electrode.\\

\noindent\textbf{Flow measurements.}
To measure the fluid flows induced by the spinner motion, we image the single-spinner setup from above using a LUX160 (Ximea) camera equipped with a zoom lens. 
The image size is $3648 \times 3402\,\mathrm{px^2}$ and the resolution is $152 \,\rm px/mm$. 
We first detect the bead in each frame using the circle detection algorithm \texttt{viscircles} implemented in MATLAB. 
The passive tracers allow us to measure the velocity field using standard particle image velocimetry (PIV). 
We use the MATLAB module PIVLab2.
Once the bead position ${\bf R}_0(t)$ and the velocity field ${\bf v}({\bf r},t)$ are obtained for each frame $t$, we perform the coordinate transformation
\begin{equation}
{\bf r}' = {\bf r} - {\bf R}_0(t).
\end{equation}
We then compute the time-averaged velocity field in the bead-centered frame,
\begin{equation}
{\bf \tilde{v}}({\bf r}') = \langle {\bf v}({\bf r}',t) \rangle_t .
\end{equation}
%


\subsubsection{Fractal geometry of the spinner clusters }

We characterize the fractal geometry of the clusters and holey fluids using a set of observables and critical exponents commonly employed to describe percolation structures \cite{saleur1987exactSI}, polymer physics \cite{duplantier1987exactSI}, and spin systems \cite{jaubert2011analysis}.\\

\noindent \textbf{Cluster analysis.} 
We consider two particles to belong to the same cluster if their centers are separated by a distance smaller than $2R+\epsilon$, with $\epsilon = 3.36\,\mathrm{mm}$. 
The choice of $\epsilon$ is  arbitrary. In practice, clusters are identified using the \texttt{dbscan} Matlab function.\\

\noindent
\textbf{Radius of gyration.} 
The radius of gyration of a cluster measures the average distance of the particles in that cluster from its center of mass. It is defined as
\begin{equation}
R_g^2 = \frac{1}{s} \sum_{p=1}^{s} 
\left( {\bf r}_p - \frac{1}{s}\sum_{p=1}^{s}{\bf r}_p \right)^2 ,
\end{equation}
where ${\bf r}_p$ is the position of particle $p$ and $s$ is the number of particles belonging to the cluster. 
For self-similar clusters, the radius of gyration scales with the cluster size as
\begin{equation}
R_g \sim s^{\nu},
\end{equation}
where $d_f=1/\nu$ is the fractal dimension of the clusters.
Values of $d_f$ close to $2$ correspond to compact aggregates approaching close packing, whereas values closer to $1$ indicate elongated or branched structures.\\

\noindent
\textbf{Cluster-size distribution and correlation function.} 
For self-similar clusters, the cluster-size distribution follows a power law
\begin{equation}
P(s) \sim s^{-\tau}.
\end{equation}


\subsubsection{Transition from gel to crystal}



\noindent {\bf Orientational order.} We quantify the orientational order of our system using the strandar bond orientational parameter $\psi_6$. For a particle $j$, it  is defined as:
\begin{equation}
\displaystyle \psi_{6,j} = \frac{1}{N_v} \sum_{k=1}^{N_v} e^{i6\theta_{jk}},
\end{equation}
where $N_v$ is the number of neighbors of particle $j$, and $\theta_{jk}$ is the angle of the vector ${\bf r}_j - {\bf r}_k$ relative to the horizontal. Thus, $|\langle \psi_{6,j} \rangle_j| = 1$ if the system exhibits perfect hexatic order. 
In practice, we compute $\psi_{6,j}$ for each particle $j$ using a Voronoi tessellation.

We also measure the correlation length $\xi_6$ of the hexatic order. The $\psi_6$ correlation function $g_6$ is defined as:
\begin{equation}
g_6(r) = \langle \psi_6(0)\psi_6^{\star}({\bf r})\rangle, 
\end{equation}
where $z^\star$ denotes the complex conjugate of $z$. We then determine $\xi_6$ via an exponential fit of $g_6(\mathbf r)$

\section{Rolling-to-hovering crossover}

\begin{figure*}
\includegraphics[width=\textwidth]{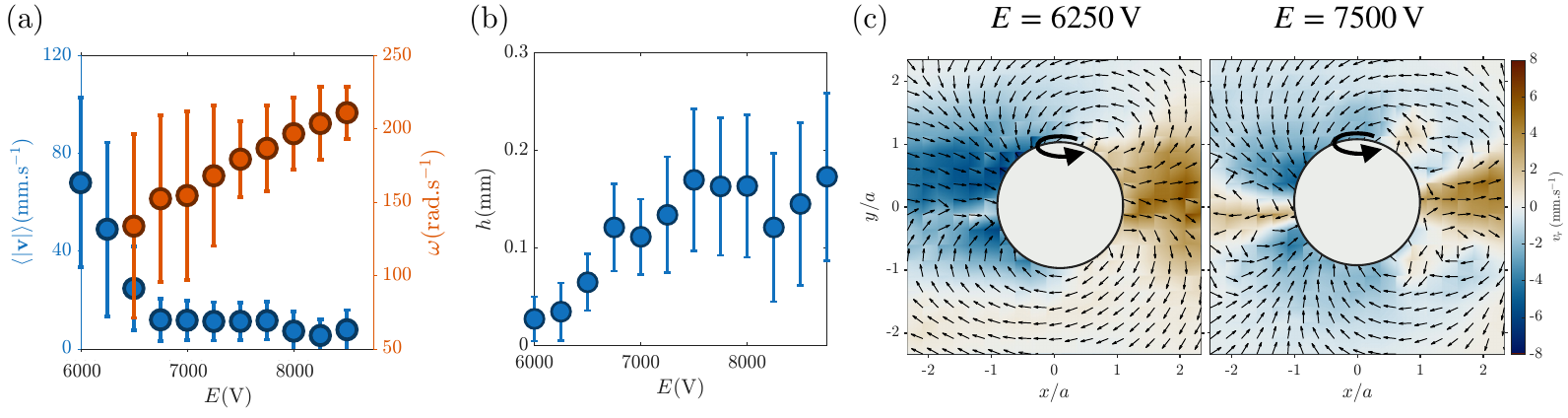}
\caption{\textbf{Rolling-to-hovering crossover.}
{\bf a.} Translational speed $v$ and scaled spinning speed $a\omega$ plotted versus the electric field $E$. 
Rolling is progressively suppressed at high field while spinning is enhanced. 
{\bf b.} Distance $h$ between the center of the spinner and the bottom electrode.
{\bf c.} Flow fields averaged over the cell thickness at $E=6250\,\rm V$ and $E=7500\,\rm V$, respectively. 
The transition from dipolar to quadrupolar circulation is clearly visible and quantified by the corresponding angular spectra shown in the bottom row.
}
\label{RollingHovering}
\end{figure*}

In this section we discuss the transition from rolling to hovering for Quincke particles operating above a flat electrode. 
This transition is not sharp but rather corresponds to a smooth crossover between the two dynamical regimes as the magnitude of the electric field increases. 
It is revealed by a set of consistent measurements summarized in Figure~\ref{RollingHovering}.

We begin by examining the particle kinematics. 
Figure~\ref{RollingHovering}a shows that the rotation speed $\omega$ increases as $E$ increases, as predicted by the Quincke theory \cite{quincke1896ueberSI}. However, the translational speed continuously decreases and eventually plateaus at a very small value, while the spinning speed keeps increasing.
Such behavior is inconsistent with pure rolling motion and contrasts with that of Quincke rollers, whose translation speed typically increases with the electric field.

This apparent contradiction is resolved by measuring the distance $h$ between the bead center and the bottom electrode (Fig.~\ref{RollingHovering}b). 
As the electric field increases, $h$ grows steadily, showing that the particles progressively detach from the surface and transition from rolling to hovering. 
Once the particle hovers above the electrode, both solid and viscous friction  with the surface are strongly reduced, which naturally explains the strong suppression of the translational velocity.

This crossover also qualitatively alters the hydrodynamic flows generated by the particle motion (Fig.~\ref{RollingHovering}c). 
When the particle rolls along the electrode, the induced flow displays the dipolar symmetry expected from mass conservation near a no-slip boundary. 
By contrast, when the particle hovers and primarily spins, the vertically averaged velocity field exhibits a pronounced quadrupolar component. 
Such flow structures are characteristic of inertial secondary flows generated by vortices at finite Reynolds numbers~\cite{shen2023,gelvan2025SI,Chen2025SI}.

Hovering has previously been attributed to dielectrophoretic forces acting on Quincke rollers~\cite{pradillo2019SI}. 
In our experiments it is additionally promoted by the inertial lift experienced by a sphere spinning close to a solid surface. 
At high electric fields, both mechanisms counteract gravity and stabilize the hovering state.

\section{Hydrodynamic torques}

We recall that the spinning axis of an isolated Quincke particle results from a spontaneous symmetry breaking~\cite{Taylor1969SI,Bricard2013SI}. 
In the geometry of our experiments, once the electric field reaches the threshold $E_Q$, rotational symmetry is spontaneously broken and the particle begins to spin about any axis lying in the plane transverse to $\mathbf E$. 
However, applying an external torque $\bm \tau$ to a Quincke particle explicitly breaks this rotational symmetry and selects the direction of rotation such that $\bm \omega \propto \bm \tau$~\cite{Bricard2013SI,Morin2018}.

\begin{figure}
 \includegraphics[width=0.9\columnwidth]{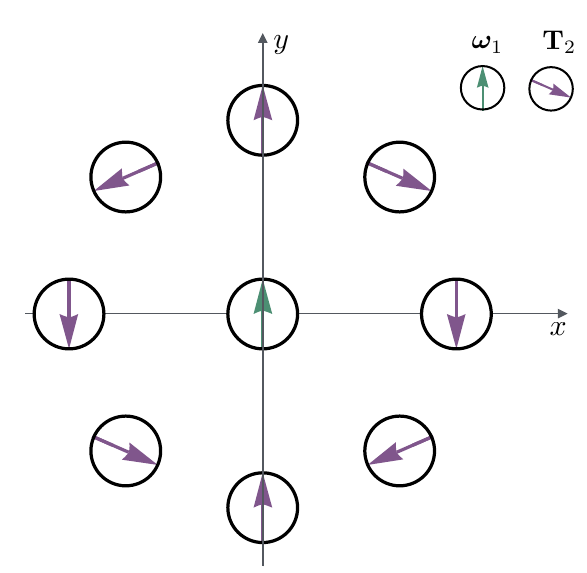}
\caption{{\bf Lubrication torque induced by a spinning particle on a neighboring sphere.}
We consider a sphere centered at the origin and rotating with angular velocity $\bm \omega_1=\omega_1\hat{\mathbf y}$ (green arrow). 
Using Eqs.~\eqref{eq:Tx} and~\eqref{eq:Ty}, we compute the torque acting on a neighboring sphere located at a distance $r$ such that $\xi=10^{-3}$. 
Purple arrows indicate the orientation of the induced torque $\mathbf T_2$ for different relative particle positions.
When the two spheres are aligned along $\bm \omega_1$, the $X^C$ contribution in Eq.~\eqref{eq:torque} dominates: $T_2^x$ vanishes and $T_2^y\propto \bm \omega_1$, corresponding to a ferromagnetic-like coupling. 
When the spheres are aligned along the direction perpendicular to $\bm \omega_1$, the $Y^C$ contribution dominates: $T_2^x$ again vanishes and $T_2^y\propto -\bm \omega_1$, corresponding to an antiferromagnetic-like coupling. 
For all other orientations, both terms contribute to the hydrodynamic torque.
}
\label{SpinnerTorques}
\end{figure}

At high packing fraction, the number of spinners in close proximity increases, viscous forces and torques become significant, and the spinning dynamics of neighboring Quincke spinners become hydrodynamically coupled.
Jeffrey and Onishi~\cite{jeffrey1984SI} computed the torque $\mathbf T_2$ generated by lubrication forces on a sphere immersed in a viscous fluid and in near contact with an identical sphere located at position $\mathbf r$ and rotating with angular velocity $\bm \omega_1$. 
The resulting torque takes the general form
\begin{equation}
\mathbf T_2=\left(X^C \mathbf P+Y^C \mathbf P^\perp\right)\cdot \bm \omega_1,
\label{eq:torque}
\end{equation}
where the projection operators are defined as $\mathbf P=\hat{\mathbf r}\hat{\mathbf r}$ and $\mathbf P^\perp=\mathbf I-\hat{\mathbf r}\hat{\mathbf r}$.
The coefficients $X^C$ and $Y^C$ quantify the axial and transverse hydrodynamic couplings between the spinners (Fig.~\ref{SpinnerTorques}a). 
The coefficient $Y^C$ can be interpreted as the viscous coupling between two gears in contact and rolling in opposite directions (an antiferromagnetic-like interaction), whereas $X^C$ represents the viscous coupling between two spinners aligned along their rotation axis and co-rotating (a ferromagnetic-like interaction).

Within the lubrication approximation these coefficients take the form
\begin{align}
X^C&={\pi\eta a}{\zeta(3)},\\
Y^C&=\frac{2}{3}\pi\eta a \log(\xi),
\end{align}
where $\xi=(r-2a)/a$ is the rescaled separation between the spheres. 
The coefficient $X^C$ remains finite and positive as $\xi\to0$, whereas $Y^C$ is negative and diverges logarithmically in this limit.

To understand the angular dependence of the pair correlations $g_{\bm \omega}$, we consider a spinner located at the origin and rotating about the $y$ axis with angular velocity $\omega_1$, and denote $\mathbf r=(r\cos\varphi,r\sin\varphi)$. 
The $x$ and $y$ components of  $\mathbf T_2$ then read
\begin{align}
T_2^x&=\omega_1(X^C-Y^C)\sin\varphi\cos\varphi,
\label{eq:Tx}
\\
T_2^y&=\omega_1\left(X^C\sin^2\varphi+Y^C\cos^2\varphi\right).
\label{eq:Ty}
\end{align}
These expressions show that when the spinners are separated along the axial direction ($\varphi=\frac{\pi}{2}$), $T_2^x=0$ and $T_2^y>0$. 
Viscous hydrodynamics therefore promotes co-rotation, consistent with Fig.~3X in the main text. 
Conversely, when the spinners are separated along the transverse direction ($\varphi=0$), $T_2^x=0$ and $T_2^y<0$, so viscous hydrodynamics favors counter-rotation, again in agreement with our experimental observations.
More generally, using Eqs.~\eqref{eq:Tx} and~\eqref{eq:Ty} we show in Fig.~\ref{SpinnerTorques}b how the hydrodynamic torque generated by spinning motion depends on the relative position of the particles.

\section{Supplementary videos}
 
\textbf{Supplementary movie 1.} Individual dynamics of a spinner at a voltage $E = 8750 \, \rm V$. 
The spinner rotates at a rate of $\sim 30 \, \rm Hz$ but barely translates. 
Spinner size : $6 \, \rm mm$ diameter. 
Video recorded at $300 \, \rm fps$, slowed down $10$ times,  and played at $30 \, \rm fps$. \newline

\textbf{Supplementary movie 2.} First part: experimental video of the cluster liquid phase ($E = 8750 \, \rm V$ and $\phi = 0.15$). 
The spinners form small clusters that constantly break and reorganize.   
Video recorded at $300 \, \rm fps$, slowed down $10$ times,  and played at $30 \, \rm fps$. 
Second part: the particles are shown with color encoding their spin. There is no spin order in this phase. Slowed down $5$ times.  \newline

\textbf{Supplementary movie 3.} First part: experimental video of the holey liquid phase ($E = 8750 \, \rm V$ and $\phi = 0.51$). 
The spinners are organized in a single system-spanning dynamical cluster.  Video recorded at $300 \, \rm fps$, slowed down $10$ times,  and played at $30 \, \rm fps$. 
Second part: the particles are shown with color encoding their spin. There is no spin order in this phase. Slowed down $5$ times.  \newline

\textbf{Supplementary movie 4.} First part: experimental video of the spin-textured crystal phase ($E = 8750 \, \rm V$ and $\phi = 0.67$). The system is phase-separated between a dense crystal and a dilute liquid-like phase. Video recorded at $300 \, \rm fps$, slowed down $10$ times,  and played at $30 \, \rm fps$. 
Second part: the particles are shown with color encoding their spin. There is a clear spin order in this phase: spins are ferromagnetically ordered in one axis of the crystal, and antiferromagnetically ordered in the other axis. Slowed down $5$ times. \newline

\textbf{Supplementary movie 5.} Experimental video of the flock phase ($E = 6000\, \rm V$ and $\phi = 0.43$). At this relativeley low electric field, the particle do not spin but roll on the bottom electrode. Video recorded at $300 \, \rm fps$, slowed down $10$ times,  and played at $30 \, \rm fps$. \newline

\textbf{Supplementary movie 6.} Experimental video of the secondary flow generated by an active spinner at an applied voltage of $E = 8500\, \rm V$. 
The centrifugal force exerted on the fluid by the high-speed rotation of the bead generates two jets along the directions perpendicular to the spinning axis. Mass conservation then implies that the expelled fluid is subsequently pumped back toward the poles of the bead. The video clearly illustrates these two basic mechanisms, which govern the angular symmetry of the secondary flows.
The primary vortex generated by the rotation of the bead is not captured by this top view and the large depth of field of our camera lens. 
Video recorded at $100 \, \rm fps$, slowed down $4$ times,  and played at $25 \, \rm fps$. 

%